\documentclass[twocolumn,showpacs,preprintnumbers,amsmath,amssymb,eqsecnum]{revtex4}
\usepackage{graphicx}
\input epsf
\usepackage[T2A]{fontenc}
\usepackage[russian]{babel}
\def\({\left(}
\def\){\right)}

\def \Z{{\mathbb Z}}
\newcommand{\beq}{\begin{equation}}
\newcommand{\eeq}{\end{equation}}
\newcommand{\bea}{\begin{eqnarray}}
\newcommand{\eea}{\end{eqnarray}}

\newcommand{\bean}{\begin{eqnarray*}}
\newcommand{\eean}{\end{eqnarray*}}
\newcommand{\bs}{\begin{subequations}}
\newcommand{\es}{\end{subequations}}

\begin{document}
\title{An Infrared Divergence in the cosmological measure theory and the anthropic reasoning}
\author{A.V. Yurov${}^{1}$, V.A. Yurov${}^{1,2}$, A.V. Astashyonok${}^{1}$, A.A. Shpilevoi${}^{1}$\\
\small ${}^{1}$Baltic Federal University of I. Kant\\
\small Department of Theoretical Physics, 236041\\
\small 14, Nevsky st., Kaliningrad, Russia\\
\small ${}^{2}$ Department of Mathematics, University of Missouri, Columbia\\
\small Columbia, 65201, USA}

%\small $*$ ГОУ ВПО Ульяновский государственный университет\\
%\small Российский  государственный  университет им. И. Канта\\
%\small 432000,  ул. Л. Толстого, 42, Ульяновск, Россия\\
%\small e-mail: sv\_chervon@rambler.ru\\
%\small $^{+}$ Кафедра теоретической физики\\
%\small Российский  государственный  университет им. И. Канта\\
%\small 236041, ул. Александра Невского, 14, Калиниград, Россия\\
%\small $^{+}$e-mail: artyom\_yurov@mail.ru }
%\date {}

\begin{abstract}

An anthropic principle has made it possible to answer the difficult question of why the observable value of cosmological constant ($\Lambda\sim 10^{-47}$ GeV${}^4$) is so disconcertingly tiny compared to predicted value of vacuum energy density $\rho_{SUSY}\sim 10^{12}$ GeV${}^4$. Unfortunately, there is a darker side to this argument; being combined with the cosmic heat death scenario, it consequently leads to another absurd prediction: the probability of randomly selected observer observing $\Lambda=0$ ends up being exactly equal to 1. We'll call this controversy an infrared divergence problem. It is shown that the IRD prediction can be avoided with the help of a singular runaway measure coupled with the calculation of relative Bayesian probabilities by the means of the {\em doomsday argument}. Moreover, it is shown that while the IRD problem occurs for the {\em prediction stage} of value of $\Lambda$, it disappears at the {\em explanatory stage} when $\Lambda$ has already been measured by the observer.

\end{abstract}


\maketitle

\section{Introduction}

Among the challenges the physics is facing nowadays, the problem of vacuum energy, a.k.a. the cosmological constant's problem is probably most troublesome of all. It's initial advent was heralded by an incomprehensibly huge discrepancy (by no less then 50 orders of magnitudes) between the predicted and observable value of cosmological constant. The furthest investigations has aggravated this already most grievous observation by adding to it a problem of fine tuning in early universe. To put it simply, the sheer magnitude of the currently observable value of $\rho_{{\rm vac}}$ implies that for some unknown reason the vacuum energy density in the early universe had to be equal to the characteristic value of $\rho_{GUT}$ (the change in the vacuum energy induced by a phase transition of a GUT scale) and, what is worse, the equality had to hold with an unbelievable accuracy of up to $10^{-107}$ -- unbelievable even more since we are talking about two distinct and (as far as we know) entirely unrelated quantities.

There were many attempts to amend the described situation, including the explanations, based on the alternative cosmological models, such as brane world models and the cosmologies with the variable speeds of light \cite{Me1}, \cite{Me2}. In the case of classical cosmology, however, the honor of explanation is widely relegated to the anthropic principle. In the well-known article \cite{Wein} it has been shown that the upper boundary for the effective cosmological constant is
$$
\Lambda_{max}<5000\Lambda_{0},
$$
where $\Lambda_{0}$ -- is an observable value. The higher values effectively prevents the observable large-scale structure from formation, and, therefore, lead to cosmologies completely devoid of life as we know it. In view of this, the aforementioned fact of fine tuning in {\em observable} universe stops being mysterious at last.

The last important step in the development of the applicable anthropic reasoning was inclusion of selection rules such as a self-sampling assumption or  mediocricity principle. It is important to note, that acceptance of any of those hypothesis, firmly grounded on statistical approach, explicitly implies the existence of a multiverse, that serves a role of a statistical ensemble. By accepting them, we are opening a road to estimating the probabilities of observing any given event $j$. Such a probability is factorizable as (cf. \cite{Adams}):

\begin{equation}
P_j\sim {\bar{P}}_jf_j,
\label{factor}
\end{equation}
where ${\bar{P}}_j$ is the concentration of j-type bubbles and $f_j$ is the anthropic factor proportional to a total amount of observers residing inside the j-type bubble.
\newline
{\bf Remark 1.} It should be pointed out that although this factorization serves as a very good approximation for many choices of cosmological measure (e.g., for the causal patch measure), there are some measures that dictates certain changes to formula. For example, the geometric cutoff measure require not $f_j$, but $f_j(t)$, with $t$ denoting the time of the bubble nucleation.

Of course, in order for (\ref{factor}) to make sense we have to devise a way to count down the observers. In what follows we'll use the idea that the existence of observers has to be equivalent to the generation of entropy production. This idea in turn follows from the fact that every observation should increase the entropy of the system and that in average the total amount of observations would be proportional to the amount of matter's entropy produced in the volume under consideration. This idea has already been successfully approbated in \cite{Bousso-2} to calculate the expected average value of cosmological constant in the observable universe.

However, despite the successes, the cosmologists has recently gotten strong indications that not everything is as simple and clear with anthropic reasoning as they believed it was. In an ironic twist, the anthropic reasoning, while solving the aforementioned problem, seems to create yet another one. In fact, it predicts the probability $P_j$ to find oneself in a region with nonzero $\Lambda$ to be exactly zero. This stunning result is based on a fact that the fraction of observations $f_j$ made in universes with $\Lambda\ne 0$ has a zero measure due to heat death befalling such universes and all their inhabitants \cite{Adams}. In other words, the anthropic reasoning asserts that with 100\% certainty we have to be in a universe with $\Lambda=0$. \footnote{This prediction has been known even before the advent of the aforementioned method - cf. for example, \cite{Barrow03}.} This discrepancy we'll call an infrared catastrophe or infrared divergence of cosmological measure. Another surprise has revealed itself in the article \cite{gonz-Elis}, in which the probabilities ${\bar{P}}_j$ were proved to be equally zero for models with $\Lambda>0$ and nonzero for the ones with $\Lambda\le 0$. These unforseen discoveries gave some authors the reasons to claim either limited applicability or even complete inadequacy of anthropic reasoning. The main purpose of this article is to provide the counterarguments to such a point of view. We will show that the zero probabilities doesn't necessarily pose a threat to validity of anthropic reasonings, although they can certainly be of use for the task of determining the limitations of the method. Our proof will be based on analysis of prediction and explanation as two distinct types of anthropic reasoning, made in article of Garriga and Vilenkin \cite{explanation}; we'll show in particular that the arguments of Garriga and Vilenkin holds true even for those cases when the predicted probabilities turn to zero.

On the other hand, some might argue that the remedy is not much better then the illness itself, as the prospect of reduction befalling the status of the anthropic principle from being the one {\em predicting} things to a mere {\em explanatory} argument truly cannot be called very promising. We'll show that this problem might actually be amended by utilizing a different rule for counting the probabilities, namely the one used in formulating the Doomsday argument~\footnote{The argument in question has seen a lot of criticism, but no proof of its fallibility has been found so far.}. The change in the rule might be attributed to indexical information pertaining to the observer and required for successful calculations regarding his/her own existence \cite{explanation}, \cite{Bostrom-index}.

Here is the main plan of the article. The second chapter will be devoted to reviewing the results leading to an infrared divergence of cosmological measure. In the next chapter these results will be shown to lead to no antimony, provided one use the anthropic reasoning to {\em explain} the obtained value of $\Lambda$ instead of {\em predicting} it. In the fourth chapter we will show that incorporation of Doomsday argument into the calculations of Bayesian probabilities effectively disposes of the infrared divergence, leaving the objections against the anthropic reasoning ungrounded and thus rectifying the problem spotted by \cite{Adams}. The a priori distribution used therein is an exponential runaway measure (singular weighting function) recently studied by Linde  and  Vanchurin in \cite{Linde-Vanchurin-2010}. The final chapter contain the final remarks and the conclusion.

\section{An infrared divergence: The core of a problem}

An advent of infrared divergence was inevitable after it was realized (cf. \cite{Bousso-2006}) that the regions with  $\Lambda=0$ are open FRW universes, with unbounded entropy and lifetime and thus containing a maximal -- infinite, as a matter of fact, -- amount of observers. Below we'll list the most important numerical estimates made in \cite{Bousso-2}, \cite{Bousso-1}.

The probability $P_{\Lambda}$ to find oneself in a universe with a given value of vacuum energy can be determined via
\begin{equation}
P_{\Lambda}\sim N(\Lambda)\bar{P}(\Lambda),
\label{1}
\end{equation}
where $\bar{P}(\Lambda)$ is an a priori probabilities distribution (the relative abundance of different values of $\Lambda$ associated with the different types of bubbles in the multiverse), and $N(\Lambda)$ is an anthropic factor proportional to the total number of observers in a given region of a Multiverse. The aforementioned number is evidently related to the star formation rate, which can in turn be estimated from the astrophysical data (cf. for example \cite{Bousso-3}), resulting in:
\begin{equation}
N(\Lambda)\sim \int^{t_{c}}_{0} \dot{n}(t)V_{c}(t)dt,
\label{2}
\end{equation}
where $\dot{n}(t)$ is the star formation rate in a comoving volume $V_{c}$ and $t_{c}$ defines a time of collapse. Obviously, for those universes whose expansion has a de Sitter-type asymptote $t_{c}=\infty$.

The behaviour of $\dot{n}(t)$ depends on the particular model of star formation \cite{Hernquist},\cite{Nagamine},\cite{Hopkins}. However, interestingly enough, the difference is not essential, since all of them predicts $\dot{n}(t)$ reaching a maximum after a couple of billions of years and then suffering a relatively fast decrease. Of course, the height and the width of the maximum depend on the cosmological constant.

If the universe has a zero curvature and the radiation is small enough, the Friedman equations become integrable and the scale factor appears to have a following time dependence:
$$
a(t)=At_{\Lambda}^{2/3}\sinh^{2/3}(3t/2t_{\Lambda}),
$$
where  $t_{\Lambda}=\sqrt{3/\Lambda}$, and $A$ -- a constant of integration. The vacuum energy's influence dominates over the dynamics from $t_{\Lambda}$. For $t>t_{\Lambda}$ the expansion rate can be correctly approximated as:
$$
a(t)\sim t_{\Lambda}^{2/3}\exp(t/t_{\Lambda}).
$$
For $t\ll t_{\Lambda}$ the scale factor changes according to a power law
$$
a(t)\sim t^{2/3}.
$$
The next step will be the determination of a measure suitable for calculating the comoving volume  $V_{c}$. In this chapter we'll use as example the causal patch cut off: the measure that is defined as an ensemble of points lying inside of a light cone in the past of a point on the future boundary of spacetime, rather popular among the cosmologists and, most importantly, consistent with formula (\ref{factor}) (cf. Remark 1).

Such a choice for $t>t_{\Lambda}$ will result with high accuracy in
$$
V_{c}(t)=\frac{4\pi}{3} \left(\int^{\infty}_{t}\frac{dt}{a(t)}\right)^{3}\sim
t_{\Lambda}\exp(-3t/t_{\Lambda}).
$$
When $\Lambda\rightarrow 0$ $V_{c}(t)$ diverges. For $t<t_{\Lambda}$ it is possible to write the expression for $V_{c}$ with accuracy up to the linear terms of $t$:
$$
V_{c}(t)\sim \left(C-3t^{1/3}\right)^{3},
$$
where $C$ is a constant, appropriate choice of which should allow for continuity of $V_{c}$ together with its first derivative at $V_{c}$. The integral, appearing in (\ref{2}) can be broken into two integrals over the intervals $0<t<t_{\Lambda}$ and $t>t_{\Lambda}$ consequently. The second integral always converges as it contains an exponentially decreasing function  $V_{c}(t)$.

The anthropic factor is inversely related to the cosmological constant. The resulting $P(\Lambda)$ will be determined by the previously chosen function $\bar{P}(\Lambda)$. If the probability $\bar{P}(\Lambda)$ increases proportionally to $\Lambda$ (at least for those $\Lambda$ that are anthropically acceptable) then at certain $\Lambda_0$ the probability to find oneself in the universe with a given value of vacuum energy would be maximal.

It is possible to convey similar analysis for negative $\Lambda$'s. The problem arises, however, when we include in our considerations the zero value of cosmological constant. In this case the casual patch volume $V_{c}\rightarrow\infty$ regardless of $t$. As a result, the anthropic factor  $N(0)\rightarrow\infty$, and we have to conclude that our universe with necessity should have had $\Lambda=0$! \footnote{Some cosmologists prefer to talk not about $\Lambda=0$ itself, as such a supersymmetric vacua might significantly alter a chemistry in the universe, substituting the electrons with the selectrons and deeming the life impossible. Instead, they refer to the infinitesimally close neighbourhood of value $\Lambda=0$, whose members, taken together, serves the role of the ``most occupied slice of multiverse''. Since we quite evidently do not belong there, the problem remains.}

This problematic result can be aggravated further by noting that any universe with a positive cosmological constant eventually turns into a dS universe. When the temperature drops to $T_{dS}\sim \Lambda^{-1/2}$ the entropy production ceases to be and the universe enters into what was quite aptly called a cosmic heat death \cite{Adams}. In other words, a dS universe can only contain a finite number of observers per unit volume, which is quite different from a universe with a zero cosmological constant. Since the functions $f_j$ in (\ref{factor}) are nothing but relative fractions of all possible observations, performed in every nook and corner of the multiverse, it immediately follows that the anthropic factor attributed to a universes with $\Lambda>0$ is $f_j=0$.
\newline
{\bf Remark 2.} The aforementioned reasoning is somewhat different from the one made in \cite{Adams}. In fact, their arguments were based on the Copernican Time Principle \cite{Adams-1}, an assumption that all moments of time in the past, present, and future of our universe should enter our considerations on equal terms. The one adopting the Principle is immediately rewarded by applicability of a self-sampling assumption for all the four-dimensional volume ~\footnote{The Principle will be closely considered in the fourth chapter.}. Now, recalling that the dS universe is itself eternal while the time period in which it can sustain life is finite, it is natural to conclude that probability to find oneself in such a universe is equally zero.
%%%%%%%%%%%%%%%%%%%%%%%%%%%%%%%%%%%%%%%%%%%%%

The problem we have stumbled upon leads us to a following question.

It is a prediction of the eternal inflation model that there exist a cosmological multiverse, containing infinite number of copies of every possible observer. What can be said about probabilities of a event in such a multiverse?

Suppose we are conducting an experiment with two possible outcomes $A$ and $B$. The probabilities for those outcomes are connected by formula
$$
\frac{P_{A}}{P_{B}}=\frac{N_{A}}{N_{B}},
$$
where $N_{A,B}$ is the amount of $A$ and $B$ outcomes in all the multiverse. The direct calculation of probabilities then becomes an insurmountable task as $N_{A,B}\rightarrow\infty$.

Getting over this difficulty requires a procedure for elimination of infinities, i.e. a suitable measure. Some possible candidates have been proposed
so far \cite{Bousso-4}-\cite{9}.

For the purpose of illustration let us consider a toy landscape with two types of universes differing by their cosmological constant.

The total number of events has a form:
\begin{equation}
\begin{array}{l}
N_{A,B}=N_{1}\int^{t_{max}}_{0}\rho^{(1)}(t)_{A,B}V_{c1}(t)dt+\\
\qquad \qquad+N_{2}\int^{t_{max}}_{0}\rho^{(2)}(t)_{A,B}V_{c2}(t)dt,
\label{3}
\end{array}
\end{equation}
where $\rho^{(1,2)}(t)_{A,B}$ is the number of $A$ ($B$) events per four-dimensional unit volume of universe 1 (2).
Ratio $P_{1,2}=N_{1,2}/(N_{1}+N_{2})$ defines the relative abundance of universe 1 or universe 2 in the theory
landscape.

In case of a universe with $\Lambda\neq 0$ the comoving volume $V_{c}$ converges to a finite value thus getting rid of infinities. The integrals in  (\ref{3}) ends up being finite and
$$
\frac{P_{A}}{P_{B}}=\frac{P_{1}n^{(1)}_{A}+P_{2}n^{(2)}_{A}}{P_{1}n^{(1)}_{B}+P_{2}n^{(2)}_{A}},
$$
where $n^{1,2}_{A,B}$ is the number of events in the single vacuum (single universe). From here it becomes obvious that in case of $\Lambda=0$ (the therminal vacuum) these numbers all go to infinity dealing a crushing blow to our calculational endeavour.

In the arguments above we have introduced an a priori probability distribution  $\bar{P}_{\Lambda}$, a fraction of universes with a particular value of a cosmological constant. However, a somewhat different approach has been put forward in \cite{explanation}. According to it, one has instead to utilize an a priori probability distribution $P_{V_{c}}(\Lambda)$, characterizing a fraction of (comoving) volume occupied by regions with a given value of
$\Lambda$. The merit of this approach is that it entirely waives the difficulties with zero cosmological problem. The equivalence of two approaches for $\Lambda\neq 0$ is indisputable and requires no additional comments. As for the case $\Lambda=0$, a simple assumption $\bar{P}(\Lambda)\neq 0$ makes sure that a fraction of comoving volume, occupied by universes with zero $\Lambda$ is equal to 1, since a
comoving volume for even one such universe is infinitely large. Vice versa, if $0<P_{V_{c}}(0)<1$, then $\bar{P}(0)=0$. It is interesting to note, that the case $\bar{P}(\Lambda)\sim \Lambda$ considered in \cite{Bousso-1} leads to similar results.

It was an infrared divergence resulting in $f=0$ that we have been discussing so far. However, the recent work of Ellis and Smolin \cite{gonz-Elis} provides us with arguments that the positive values of  $\Lambda$ might as well correspond to a zero value of a first multiple in (\ref{factor}).

To make it clear, one need to take into account that the landscape consists of an infinite set of discrete vacua, which are nothing but different   possible (metastable) ground states of string theory. It was shown in \cite{12-12}   that these vacua include $N_+\sim 10^{500}$ flux compactifications which  appear compatible with $(3+1)$ noncompact dimensions and a {\em positive} cosmological constant. On the other hand, estimation of the number of allowed vacua with a negative cosmological constant anthropically results in $N_-=\infty$ \cite{13-13}, \cite{14-14}. Using the weak anthropic principle, one get rather a discouraging outcome:  the relative probability for an observer to find him/herself in a universe filled with a positive cosmological constant is {\em zero}.  More precisely, for a given anthropic limit  $\Lambda_+>0$ and for any positive $\epsilon \ll \Lambda_+$ we have the probability $P(\Lambda > \epsilon)=0$ \cite{gonz-Elis}. This is true because there should be infinitely more universes with negative $\Lambda$ than universes with strictly positive cosmological terms.

However, it should be remembered that a unverse with negative  $\Lambda$ inevitably reaches a contraction phase that ends up in a final singularity. The total amount of possible observations in such a universe is necessarily bounded and as such can't compare with the numbers, reached in the unverse with nonexistent (zero) cosmological constant. After the dust settled we are still one-on-one with the infrared divergence problem.

\section{explanation vs. prediction}

The challenge of the infrared divergence did not not remain unanswered. The possible solutions, provided by different authors, can be arranged into the following groups:

(1) There exists a physical principle, unknown as yet, imposing the bounds on the function $\bar{P}_{\Lambda}$, so that $\bar{P}=0$ when $\Lambda=0$. For example, it might be that the vacua of nonpositive energy density simply doesn't exist.

Unfortunately, it can't be true: as follows from the analysis of Dyson, Kleban, and Susskind \cite{DKS-2002}, the real landscape has to contain vacua with nonpositive cosmological constant. The price for alternative would be a violation of both unitarity and ergodicity. Indeed, the string theory landscape is expected to contain valleys with negative cosmological constant as well as supersymmetric regions with vanishing vacuum energy \cite{KKLT-2003}. The last ones are the seeds of an infrared divergence.

(2) It is possible that the finite lifespan of metastable phases, once included into consideration, might significantly alter our conclusions.

However, the estimates for a typical lifespan $\tau$ of metastable vacua, made within a framework of the string theory \cite{KKLT-2003}, \cite{Susskind}, agree on $\tau\sim{\rm e}^{0.5\times 10^{123}}$ years. And even though some models allow to lower this number to ${\rm e}^{ 10^{19}}$ or even ${\rm e}^{10^{9}}$ years (the later is predicted by the model containing the KPV instantons \cite{Verlinde}), this can hardly be called a satisfactory solution. (\ref{2}) requires from a vacuum with $\Lambda=0$ to be abnormally short-lived compared to all other vacua. However, it has been shown in \cite{Susskind} that de Sitter minima are never stable. After a series of tunneling events they eventually end in a terminal vacua with exactly zero or negative cosmological constant.

(3) The arguments based on the anthropic reasoning might be deficient, requiring some additional assumptions. This possibility will be explored in the next chapter.

In this chapter, however, we will show that the problems uncovered above doesn't necessarily lead to drastic changes in the anthropic reasoning or the visions of cosmological multiverse. In fact, it appears that in trying to interpret the zero probability for the observable universe most of the cosmologists are making two mistakes at once.

The first mistake is the wrong interpretation of zero probabilities. Suppose a certain event $A$ has a zero probability: $p(A)=0$. Does it mean that $A$ cannot happen? The answer is: not necessarily. Those events that cannot happen are called impossible. The probability of an impossible event is zero ad definition, but the opposite is simply not true. The probability to hit a direct center of a target is zero, but is not impossible. The probability to guess number 42 out of all possible real numbers is also zero, but it can happen nevertheless. And, finally, the probability to find ourselves at one particular point in infinitely large universe is also zero, yet we are here!

Getting back to our discussion, we have seen that, according to Ellis and Smolin, an application of the anthropic reasoning to a randomly chosen observer results in a zero probability to find one in the universe with a positive vacuum energy. However, such event cannot be branded as impossible, and as such, the difference between the Ellis and Smolin prediction and the observational, as big as they are, by no means are paradoxical.

Of course, this answer immediately rises another question: isn't it true then that positiveness of the cosmological constant in observable universe is an assured proof of our {\em atypicality}? Because if the answer is {\em yes}, we have neither right no any reason to use the anthropic reasoning for any estimates we have used it for!

Fortunately for us, the answer to this question is {\em no}. Our seeming atypicality is but an offspring of a second mistake that is sadly characteristic of many articles on the usage of anthropic principle in cosmology. The one very clear and concise explanation of the mistake together with the way to avoid it has been presented in the article by Vilenkin and Garriga \cite{explanation}, which has also provided for a solution of a Hartle an Hawking ``human vs. jupiterian'' paradox  \cite{Typical?}. The goal of this chapter is to show that the reasonings of Vilenkin and Garriga remain applicable even for the outcome whose probability is exactly zero. It is actually easy to prove that the calculations of \cite{explanation} are still valid for infinite amount of ``jupiterians'' and, as a consequence, the Vilenkin-Garriga method effectively disposes of the infrared divergence problem. However, such an approach ends up being extremely cumbersome to manage. Instead, we will consider an Ellis and Smolin problem, different from the one of Hartle and Srednicki, and will show that the method of Vilenkin-Garriga does good here as well.

To put it simply, the multiverse allows for two types of conditional probabilities' calculations. The first one is called the prediction, the second one -- explanation. The difference between the probability distribution ascribed to these different types of reasoning stems from the subtle difference existing between the observers employing them. According to Vilenkin and Garriga, only the observers with similar informational content can be assigned to the same equivalence class. Suppose we don't known the sign and value of $\Lambda$ and are about to conduct an experiment that will discern them for us. We should understand that there is an infinite number of observers living in the multiverse, who, while not being similar to us, are in the same boat nevertheless and are about to embark on the same astronomical endeavor without knowing what it would yield. Assuming other factors to be insignificant for our intents (we trace over all information not correlated with the outcome of the experiment) we conclude that altogether we belong to class $C(\Lambda=?)$. For the sake of simplicity, we'll divide all possible values of $\Lambda$ into small intervals  $\Delta\Lambda_i$ and assume that the universe with $\Lambda\sim\Delta\Lambda_i$ contains $N_i$ observers. This implies that the probability of observing $\Delta\Lambda_i$ is proportional to $N_i$. Next, suppose there exist such a $J$ that $N_J\gg N_i$ for all $i\ne J$. Then the majority of observers would discover that $\Lambda\sim\Delta\Lambda_J$. But claiming that they are the {\em only} type of observers possible would just be absurd. After all, for every $\Lambda\sim\Delta\Lambda_i$ there is {\em someone} observing it!

Next, suppose the same multiversal cosmologists of class $C(\Lambda=?)$ has calculated a probability to find oneself in a universe with $\Lambda\sim\Delta\Lambda_0$ and it happen to be zero. Since this is not an impossible observation, somebody in a multiverse shall make it. What should  be their conclusion then? Should they deduce that the anthropic reasonings and the very idea of the multiverse are plain wrong? No. In order to interpret the observational results via the Bayes theorem the observers would have to take these very results into account, thus, they would have to explain, not predict.

Suppose all members of class $C(\Lambda=?)$ ponder the two mutually exclusive theories. The first one is the multiverse theory $T(M)$, and it is its prediction that there exist a multitude of universes with various values of vacuum energy $\Lambda$. Theory $T(S)$, on the contrary, insists on existence of just one universe having a certain defined $\Lambda_0$, with neither sign nor value of $\Lambda_0$ known to the observers yet. Every member of the group utilizes his or her information (the lack of it, to be certain) about $\Lambda_0$ for calculation of the likelihood measure of $T(M)$ and $T(S)$ by the Bayes formula. Of course, being the members of same group, they will come to exactly similar probabilities: $Prob(M)=Pr(M)$ и $Prob(S)=Pr(S)$. Let us also assume they all are also aware of the problem of Ellis and Smolin (or their multiversal analogues). Each one of them reasons that validity of theory $T(M)$ implies them existing in a universe with negative $\Lambda$ with 100\% certainty, because, after all, these are the universes predominating the multiverse (compare $N_-=\infty$ with $N_+<\infty$!). Now let us consider one particular astronomer $A$. The Bayesian probabilities of theories $T(M)$ and $T(S)$ as calculated by $A$ under condition of her existence are
\begin{equation}
P(T(S)|A)=\frac{Pr(S)}{1+N_-+N_+},
\label{Bs-S}
\end{equation}
\begin{equation}
P(T(M)|A)=\frac{Pr(M)(N_-+N_+)}{1+N_-+N_+}.
\label{Bs-M}
\end{equation}
Taking the limit $N_-=\infty$ $A$ gets $P(T(S)|A)=0$ and $P(T(M)|A)=Pr(M)$, which also means that $Pr(M)\sim 1$. In other words, it seems as if the theory  $T(M)$ is a priory correct, no proof required. This, of course, is wrong. The validity {\em has} to be checked through the experiment. In order to do it $A$ has to extract the predictions from both theories and compare the results with observations. The predictions regarding the vacuum energy density are also made by the Bayesian formulas:
\begin{equation}
P(A\,\,{\rm see}\,\, \Lambda>0|T(M))=\frac{N_+}{N_-+N_+}\to 0~{{\rm at}\,\,N_-\to \infty},
\label{Posit}
\end{equation}
\begin{equation}
P(A\,\,{\rm see}\,\, \Lambda<0|T(M))=\frac{N_-}{N_-+N_+} \to 1~{{\rm at}\,\,N_-\to \infty},.
\label{negat}
\end{equation}
Thus, it indeed looks like the probability to observe the negative cosmological constant in framework of $T(M)$ theory should be equal to 1. So, it is very easy to understand the embarrassment of $A$ when the test (performed by the multiversal analogues of COBE and WMAP) shows a seemingly impossible: that $\Lambda=\Lambda_0>0$ (cf. (\ref{Posit}), (\ref{negat})). It would be hard for $A$ to shake a feeling that this result is nothing short of a stab in the back of an anthropic reasoning and theory $T(M)$ -- just what Ellis and Smolin claims it to be. Indeed, taking the limit $N_-\to \infty$, $A$ arrives to formulas
\begin{equation}
\begin{array}{l}
P(A\,\,{\rm see}\,\, \Lambda_0>0|T(M))=(N_-+N_+)^{-1}\to 0,\\
\\
P(A\,\,{\rm see}\,\, \Lambda_0>0|T(S))=1,
\label{oh}
\end{array}
\end{equation}
and with the sad heart concludes that $T(S)$ is valid, and $T(M)$ is not.

But $A$ is wrong. The knowledge of $\Lambda$ changes the Bayesian formulas. The validity of $T(M)$ should now be calculated as:
\begin{equation}
\begin{array}{l}
\displaystyle{P(T(M)|A;A\,{\rm see}\, \Lambda_0)=\frac{P(T(M)|A)P_M}
{P(T(M)|A) P_M+P(T(S)|A) P_S}},\\
\displaystyle{P_M=P(A\,{\rm see}\, \Lambda_0|T(M))},\\
\displaystyle{P_S=P(A\,{\rm see}\, \Lambda_0|T(S))}.
\label{new-bas}
\end{array}
\end{equation}
Substituting (\ref{oh}), (\ref{Bs-S}) and (\ref{Bs-M}) into (\ref{new-bas}) results in
\begin{equation}
 P(T(M)|A;A\,{\rm see}\, \Lambda_0)=\frac{Pr(M)}{Pr(M)+Pr(S)}
 \label{itog-M}
 \end{equation}
and
\begin{equation}
P(T(S)|A;A\,{\rm see}\, \Lambda_0)=\frac{Pr(S)}{Pr(M)+Pr(S)}.
\label{itog-S}
\end{equation}
Therefore, the conditional probabilities for theories $T(M)$ and $T(S)$, as should be estimated by $A$ {\em after} the measurements of $\Lambda$ will be proportional to initial validity estimates $Pr(M)$ and $Pr(S)$:
\begin{equation}
 P(T(M)|A;A\,{\rm see}\, \Lambda_0)+ P(T(S)|A;A\,{\rm see}\, \Lambda_0)=1,
\label{sum1}
\end{equation}
\begin{equation}
 \frac{P(T(M)|A;A\,{\rm see}\, \Lambda_0)}{P(T(S)|A;A\,{\rm see}\, \Lambda_0)}=\frac{Pr(M)}{Pr(S)},
\label{to-je}
\end{equation}
in exact correspondence with \cite{explanation}.

\section{Exponential runaway measure and Doomsday argument vs. the infrared divergence}

The main intention behind the derivations of the previous chapter was an a priori eradication of the infrared divergence problem. Unfortunately, the method has its flaws. The most troublesome one is the fact that by resorting to such a method we leave, immediately and quite voluntarily, the field of falsifiable hypotheses, -- i.e. the multiverse hypothesis appears to be virtually unrefutable! -- and this is most definitely not a kind of outcome one can feel comfortable about. Moreover, by accepting the explanatory limitations of the anthropic reasoning, we are losing the most powerful tool in the store of any physicist: the generation of viable predictions fully testable via the experimental means. To sum it all up, the choice before us is quite clear. If we are not put off by the aforementioned limitations and are ready to accept them, we would be able to terminate the infrared divergence problem. If we are not, we'd have to invent something else.

In our opinion, in order to set the anthropic predictions right one should take a second look on the Bayes probabilities. The article \cite{explanation} has made it very clear that the meaningful probabilistic estimates are strongly connected to both the informational content and the indexical information of the observers. Two observers with identical observational data but different indexical information would come to completely different probabilistic estimates for the similar events. From this point of view, it is rational to assume that the occurrence of zero probabilities is but an artifact of our inability to account for an indexical information. Of course, we have no widely accepted criterions allowing us to make a best pick of different approaches to such a task, but then again, would it not be reasonable to choose exactly those eradicating the infrared divergence?.. And, once such a method would be found, one could start working on its justification via the indexical information, and not the other way around. This type of reasoning may seem somewhat speculative, but it appears to be rather promising. After all, it is not entirely new. The similar argument has been introduced before as an attempt to annulate the Boltzmann brain's domination problem \cite{Don} by restricting the attention only to those cosmological measures that don't allow for dominance of Boltzmann brains over the classical ordered observers \cite{BBB1} -- \cite{BBB7}.

So, how should the probabilities be calculated? According to our hypothesis, the probability to find oneself in a universe with a given value of $\Lambda$ should include a conditional probability to perform an $M_0$'s observation, where the observations are being calculated for the whole duration of the universe's existence, i.e. for 13.7 Gyr. The quantity $M_0$ is not explicitly known, of course, but it can be estimated by knowing a contemporary increment of the matter's entropy $\Delta S_0$ corresponding to a single observation \cite{Bousso-2}. In this context, the Bayes formula changes from (\ref{factor}) to:
\begin{equation}
P(\Lambda|\Delta S_0)\sim {\bar P}(\Lambda) P(\Delta S_0|\Lambda),
\label{doom}
\end{equation}
where $P(\Delta S_0|\Lambda)$ is the conditional probability of observation yielding a given cosmological value, expressed as an entropy increment.
\newline
{\bf Remark 3.} An alternative way of defining $M_0$ would be counting all stars inside of a Hubble volume with regards to to their birth time, and assigning $M_0$ to the Sun's corresponding number. However, the ``entropy'' definition is still preferable as the one explicitly related to a number of observation conveyed.

The (\ref{doom}) is intimately connected to the famous Doomsday Argument \cite{doomsday1} -- \cite{doomsday4}. The validity of this argument has been a subject of many fierce discussions. Some researchers provide the sharp criticism of the Doomsday Argument (cf. \cite{Olum1}, \cite{Korb}),  while others strongly argue in its defense (cf. \cite{Kiking}, \cite{Bs-vs-Olam}). In fact, the authors of this very article are also split in their opinions regarding the Argument ~\footnote{Two of us -- AY and VY tend to consider it correct, while AA regards it as ``probably wrong''.}. However, within the context of this article we'll accept it so it would be possible to apply (\ref{doom}) for calculating the probability of observing a given value of cosmological constant.

The first important thing would be defining an a priori distribution ${\bar P}(\Lambda)$. For this purpose we'll make use of the singular weighting function $\omega(\Lambda)$, recently studied by Linde and Vanchurin \cite{Linde-Vanchurin-2010}. This function leads to the exponentially strong runaway regimes (e.g. the Q-catastrophe \cite{Q-cat}) and can be chosen in a number of different fashions, say, as the Hawking measure
$\omega_{_H}(\Lambda)\sim \exp(24\pi^2/\Lambda)$ \cite{Hawking} or as the double exponential expression in baby universe theory $\omega_{_{BU}}(\Lambda)\sim \exp(\omega_{_H}(\Lambda))$ \cite{BU}. However, the best choice for our purposes would be the weighting function, obtained by counting the observable universes \cite{ManyUniv}:
\begin{equation}
\omega(\Lambda)\sim \exp\left(\frac{H_{_I}^{3/2}}{|\Lambda|^{3/4}}\right),
\label{ManyUniv}
\end{equation}
where $H_{_I}$ accounts for the Hubble constant at the end of the slow-roll inflation. We'd like to point out that this quantity is actually the estimate of the maximal number of different locally distinguishable classical geometries.

Granted, all these measures contain an essential singularity at $\Lambda=0$ and are subject to infrared divergence to even greater extent then the ``regular'' measures (cf. \cite{Linde-Vanchurin-2010}). Despite this, such ``pathological'' measures might actually be of great interest to cosmologists, because, being combined with the string theory they might potentially lead to isolation of a single vacuum state \cite{Linde-Vanchurin-2010}! Bearing this prospect in mind Linde and Vanchurin hypothesized that the allowed values of  $\Lambda$ form a discrete spectrum that does not include point $\Lambda=0$. If this assumption is correct, any of the three aforementioned measures can be used for calculation of the average values. Unfortunately, no real reasons for exclusion of  $\Lambda=0$ (besides the obvious fact that the observable $\Lambda$ IS positive) has been so far presented. Thus, in our further considerations we'll utilize the distribution (\ref{ManyUniv}) for the role of ${\bar P}(\Lambda)$ in (\ref{doom}) but will consider  $\Lambda$ going through a continuous spectrum of values that also includes the point $\Lambda=0$. Then the probability:
\begin{equation}
\displaystyle{
P(\Lambda_1<\Lambda<\Lambda_2|\Delta S_0)=\frac{\int_{\Lambda_1}^{\Lambda_2}\omega(\Lambda)P(\Delta S_0|\Lambda) d\Lambda}
{\int_{\Lambda_{min}}^{\Lambda_{max}}\omega(\Lambda)P(\Delta S_0|\Lambda) d\Lambda},
}
\label{doom-1}
\end{equation}
where the interval $\Lambda_{min}<\Lambda<\Lambda_{max}$ defines the anthropic bounds: $\Lambda_{max}$  is an upper Weinberg limit and $\Lambda_{min}=-\rho_{M_0}$, where $\rho_{M_0}\sim (10^{-3} \,\,{\rm eV})^4$ is the present density of matter.

Next, we shall figure out what $P(\Delta S_0|\Lambda)$ should look like. As follows from \cite{Adams}, the dS universe allows for just a finite number of observations, since at temperature $T_{dS}\sim \Lambda^{-1/2}$ the entropy production ceases and the entropy level stabilizes at $S_{dS}(\Lambda)=24\pi^2/\Lambda$. Next, since the observer might exist in any of  $\omega(\Lambda)$ universes with a given $\Lambda$, the total entropy of multiverse should be accounted for by taking a product of $S_{dS}(\Lambda)$ and $\omega(\Lambda)$. By the Copernican Time Principle no preference to any moment of time should exist, hence $P(\Delta S_0|\Lambda)=\Delta S_0/(\omega(\Lambda)S_{dS}(\Lambda))$, i.e. the fraction of the matter's entropy produced by observers of a particular universe to the total entropy amassed within the Multiverse by the universes with the particular $\Lambda$. Substituting it into (\ref{doom-1}) and taking the integrals one gets:
\begin{equation}
\displaystyle{
P(\Lambda_1<\Lambda<\Lambda_2|\Delta S_0)=\frac{\Lambda_2^2-\Lambda_1^2}{\Lambda_{max}^2-\Lambda_{min}^2}.
}
\label{otvet}
\end{equation}
If one assumes that $\Lambda_2-\Lambda_1=\delta\Lambda\ll \Lambda{1,2}$  and that $\delta\Lambda$ is fixed, one will find it very useful to utilize the probability $P_{\delta\Lambda}(\Lambda>\Lambda_1)$ with a fixed value $\Lambda_1$. For two different values $\Lambda_1$ and $\Lambda_2$ there exist a relationship:
$$
\frac{P_{\delta\Lambda}(\Lambda>\Lambda_1)}{P_{\delta\Lambda}(\Lambda>\Lambda_2)}=\frac{\Lambda_1}{\Lambda_2},
$$
in other words, we end up with a classical distribution $P_{\delta\Lambda}(\Lambda>\Lambda_1)\sim \Lambda$ and with the probability density
$\rho_{\delta\Lambda}(\Lambda>\Lambda_1)={\rm const}$. This quantity, (naively) used as an a priory distribution, leads to all the well-known results, including the well-fitting predictions regarding the observable value of cosmological constant, free from the infrared divergence problem.

\section{Conclusion}

The $\Lambda$ problem has seen a lot of attention recently, including the exotic views based on the alternative cosmological models like the  cosmologies with the variable speeds of light \cite{Me2} and brane world models \cite{Me1} (considerations of the last case made possible via construction of exact solutions having small cosmological constants, performed with the aid of Darboux transformations \cite{Me1}, \cite{BLP}). However, in this article we develop what can be called an ``anthropic approach'' instead. The article contains two main results:
\newline

A. It is shown that the zero probabilities pose no threat to the multiverse models provided one uses not a ``predictory'' but ``explanatory'' Bayes formulas.
\newline

B. It is shown that the infrared divergence problem is avoidable in the scheme, utilizing the following ingredients:

1. The probabilities are calculated according to the Doomsday Argument algorithm;

2. The a priori probabilities are defined via the exponential runaway Linde-Vanchurin measure (the singular weighting function).

3. The total entropy of the multiverse should be necessarily accounted for.

We'd like to conclude with two important remarks. First, the Doomsday algorithm should only be incorporated into the estimations of the cosmological constant value, since it is this quantity that defines the total entropy and is directly connected to the process of observation. For other parameters the algorithm of \cite{Linde-Vanchurin-2010} should be used. In other words, the suggested method doesn't corrupt the main idea of Linde and Vanchurin and allows to (i) dissolve the infrared divergence problem and (ii) validate the well-known effective distributions, that works so well for estimation of predicted $\Lambda$. Second, we have utilized the Doomsday Argument which itself is quite controversial in the eyes of many scientists. For this article we did not intend to start a discussion regarding all the ``pros'' and ``cons'' of the argument, but have merely made use of it as a method, that ends up extremely helpful in the cosmological measure theory and whose faultiness (if it exists) has not been proven (yet?). This, of course, doesn't mean that the Doomsday Argument IS correct. It merely means that if in the future the Doomsday Argument would be substantiated with high certainty, the phenomenological results of this article would acquire a very strong backup, and so would be the singular distributions of Linde-Vanchurin.

\section{Appendix. Behind the shadow of cosmological constant}

The underlying assumption behind every calculation made in this article was that the peculiar behaviour of observable universe is a direct manifestation of a non-zero vacuum energy. While this assumption is accepted by a majority of cosmologists, it is nevertheless possible that the real explanation hides in a different venue of thought. Among the different -- and quite exotic -- hypotheses regarding this matter, there is a rather unusual one we believe to be of particular interest. To be precise, it has been suggested in articles \cite{Odintsov-1}, \cite{Odintsov} that the observable cosmological effects might be attributed to a very specific dynamics of the universe, showing up on our readings as an ``effective'' $\Lambda$, while in fact there might be no ``real'' $\Lambda$ involved at all.

For example, it is a known fact that some models of modified gravity may mimic the universe with (multiple, as a matter of fact) cosmological constants (cf. \cite{Odintsov-2}). In \cite{Odintsov-1} the general scheme for modified f(R)-gravity reconstruction from any realistic FRW cosmology has been formulated, thus allowing to formulate several versions of modified gravity where the following sequence of cosmological epochs occurs: matter dominated phase (with or without usual matter), transition from decceleration to acceleration, accelerating epoch consistent with recent WMAP data, FCDM cosmology without cosmological constant. Such models can be viewed as some evidence in favor of f(R)-gravity.

Quite different approach is given in \cite{Odintsov}. Its idea goes like this. First we are reminded that the standard Einstein-Friedman equations for a flat universe might be rewritten as:

\begin{equation}
 \rho(t)=\frac{3H(t)^{2}}{\kappa^2},\quad
p(t)=-\frac{3H(t)^{2} +2H'(t)}{\kappa^2},
\label{app-1}
\end{equation}
where $p$ and $\rho$ are pressure and density, $\kappa^2=8\pi G$ and $H(t)$ is the Hubble constant $H(t)=\dot{a}/a$. This system of differential equations is open and requires some additional conditions to be solved successfully. Usually the role of such a condition is played by an equation of state, establishing the relationship between the $\rho$ and $p$. However, as follows from \ref{app-1}, it is possible to define them independently, provided one knows the exact form of function $H(t)$. As an additional bonus, we can immediately see that for any $t_0$ such that $H'(t_0)=(dH/dt)_{t=t_0}=0$ we get the relationship:
$$
p(t_0)=-\rho(t_0),
$$
which is characteristic of a universe filled with the vacuum energy. In essence, the observer watching the skies at $t=t_{0}$ will be forced to conclude that his universe contains the nonzero cosmological constant $\Lambda=3H^{2}(t_{0})$.

Of course, there is no reason for a function $H'(t)$ to have a single zero. Instead, it might contain a formidable array of zeroes, each one of them corresponding to a particular moment of time and to different observable values of ``vacuum energy''. We have therefore arrived to a classical analogue of a ``landscape'', where different space-time regions of a universe correspond to different values of ``cosmological constant''.

Among those space-time patches there exist a few with dynamics being critically similar to the one existing in the de Sitter universe. But what should be the probability to observe such a dynamics? The answer would depend on the given form of function $H(t)$. For example, for
$$
H(t)=g(h(t)-\sin h(t)),
$$
with $g$ and $h$ being some differentiable functions, its derivative turns to zero at $t=t_{n}$, with $h(t_{n})=2\pi n$ and $n \in \Z$. The effective vacuum energy therefore takes the values $\Lambda=3(g(2\pi n))^2$. Furthermore, the dynamics is close to a de Sitter one in the $\delta t_{n}$ neighbourhood of $t=t_{n}$ if the following relationship is satisfied:

\begin{equation}
\left|\frac{2\dot{H}}{3H^2}\right|=\left|\frac{2g'(h(t)-\sin h(t))(1-\cosh(t))h'(t)}{3(g(h(t)-\sin h(t))^2)}\right|\ll1.
\end{equation}

Now, the fact that we do observe such a dynamics in contemporary universe clearly leads to certain restrictions imposed on functions $g(t)$, $h(t)$. The most promising choice of function $h(t)$ appears to be a function possessing a horizontal asymptote. For example, function $h(t)=2\pi N\tanh(\omega t)$, where $\omega$ is an arbitrary constant, approaches $2\pi N$ as $t \to 0$, which means that the later stages of the universe's dynamics would be associated with an effective vacuum energy $\Lambda=3g^2(2\pi N)$ (which can actually be equal to zero -- hence, the smallness of the ``observable'' cosmological constant!). Moreover, the monotonicity of chosen $h(t)$ implies that for $N>1$ there shall exist at least $N-1$ more zeroes of function $H(t)$ that correspond to effective vacuum energies $\Lambda=3g^2(2\pi n)$, $n<N$. Therefore, we have a cosmic ``landscape'' with a plethora of various values of observable ``cosmological constants''.

Of course, the question one shall ask here is whether there exist a criterion for the functions like $g$, allowing to derive them instead of simply contriving. This problem might be somewhat simplified if one takes into account a scalar field; the problem of choice of function $H(t)$ then reduces to a problem of appropriate scalar field potential.

%%%%%%%%%%%%%%%%%%%%%%%%%%%%%%%%%%%%%%%%%%%%%%%%%%%%%%%%%

\end{document}